\documentstyle[psfig,12pt]{article}
\newcommand{\Od}{{\cal O}}

\newcommand{\PRD}[1]{{\em Phys.\ Rev.\ }{\bf D#1}}
\newcommand{\PL}[1]{{\em Phys.\ Lett.\ }{\bf B#1}}
\begin{document}
\title{{\bf Chiral symmetry and the pion gas virial 
expansion}\footnote{Research supported
by the Department of Energy under contract DE-AC03-76SF00515}}
\author{ A. Dobado$^{1}$\\
{\small \em Departamento de F\'{\i}sica Te\'orica.}\\
{\small \em Universidad Complutense, 28040 Madrid. Spain.}\\and\\
J. R. Pel\'aez$^{2}$\\
{\small \em Stanford Linear Accelerator Center}\\
{\small \em Stanford University, Stanford, California 94309}}
\date{June 1998}
\maketitle
\footnotetext[1]{E-mail:dobado@eucmax.sim.ucm.es}
\footnotetext[2]{On leave of Absence from Departamento 
de F\'{\i}sica Te\'orica.\\
Universidad Complutense. 28040 Madrid. Spain.\\
E-mail:pelaez@slac.stanford.edu}
\begin{abstract}
In this work we study the thermodynamic properties of the pion gas
starting from the realistic elastic scattering phase-shifts obtained from
Chiral Perturbation Theory and using the virial expansion. In particular
we study the state equation and the behavior of the quark condensate 
as a 
function of the temperature and the density.
\end{abstract}
\hspace*{1cm}

\vskip .5cm

\hspace*{.4cm} SLAC-PUB-7857

\hspace*{.4cm} hep-ph/9806416

\newpage

\section{Introduction}

One of the most important aspects of hadronic 
dynamics at low energies is spontaneous chiral symmetry breaking.
Indeed, when considering just the two lightest quarks as massless,
there is a spontaneous breaking of  $SU(2)_L \times SU(2)_R$
down to $SU(2)_{L+R}$ (isospin).
  As a consequence, the Nambu-Goldstone 
bosons associated with that breaking, namely the
pions, become the relevant degrees of freedom at low energies.
In the last years  a general formalism 
called Chiral Perturbation Theory (ChPT) has been developed \cite{We,GaLe}, 
which exploits the chiral symmetry constraints, providing 
a phenomenological description of low-energy pion dynamics, 
organized 
as an expansion in the pion external momenta and masses
(the latter appear due to the small explicit chiral symmetry breaking 
caused by quark masses). By using ChPT it is possible 
to describe at low energies 
a large number of processes, such as pion scattering,  in 
terms of a small set of parameters which can be fixed from 
experimental data. 

At high temperatures it is expected that chiral symmetry will 
be restored at some critical temperature, $T_c$, typically around a 
few hundred MeV. Nevertheless, below $T_c$ the main 
excitations of the hadronic medium would be long wavelength pions or, 
in other 
words, one should deal with a hot pionic gas  
\cite{SHU}. As is well known, finite temperature effects 
can be included in Quantum Field Theory (QFT) in different ways 
(see \cite{LAND} and  references therein). 
In particular this is the case in ChPT, where the imaginary time 
formalism
was used to calculate the  pion gas free 
energy  as a power expansion in the temperature over the pion 
decay constant, $F$, up to $\Od(T^6/F^6)$ \cite{GELE}. As a matter of
fact, the thermodynamics of the pion gas has also been the subject
of intense work using finite temperature QCD sum rules \cite{QCDSR}
and other alternative approaches \cite{other}.
Among other issues, some of these
calculations made it possible to obtain some $T_c$ estimates.
 In this work we will 
consider the interacting pionic gas, but using a different method which can 
be applied at higher temperatures and including 
density effects as well. Those effects are well known to be  
relevant in the study of the behavior of hadronic matter.
In particular, once they are included in the pion gas 
state equation they modify the critical temperature.  

The approach that we will follow here relies on the use of the 
relativistic virial expansion developed a long time ago \cite{DAS}. 
Such an expansion will allow us to obtain the relevant thermal 
functions in terms of the $S$ matrix, so that it is 
not necessary to go through the technicalities of finite 
temperature QFT. As a matter of fact, it is even possible to 
start from an $S$ 
matrix obtained directly from experiment, without using any 
QFT. The other key point in our approach is based on the unitarity 
of the $S$ matrix. The standard way to consider density in 
statistical physics is by introducing a grand canonical ensemble, 
whose chemical potential should be coupled to some 
conserved quantum number. In the case of the pion gas, the only 
conserved quantities are the electric charge and the baryonic number. 
In particular, the total number of pions is not conserved by 
strong interactions. However, from pion scattering data it 
is well known that  elastic unitarity is satisfied up to energies 
of about 1 GeV (where the inelastic $K \bar{K}$ channel opens up,
since the four pion channel contribution is negligible at those energies). 
Therefore, if one is only interested in  
low energy pion dynamics or low pion gas temperatures, it 
is sensible to assume that the pion number is approximately 
conserved by strong interactions. In this way it is possible 
to introduce the corresponding chemical potential and a grand 
canonical ensemble. Thus the pion density, or the number of 
pions per unit of volume, becomes a meaningful physical concept, 
at least for temperatures  well below 1 GeV.

By using the approach described above we will obtain the pion
gas state equation for low densities and temperatures below 
$T_c$. Moreover, for a given density, we will estimate this critical 
temperature as that where the quark condensate 
vanishes. The plan of the paper is as follows. 
In Sec.2 we introduce the details of the virial expansion and 
discuss in detail the free gas case, including 
the possibility of Bose-Einstein condensation.
In addition, the comparison with the free 
gas will provide an estimate of the range of applicability of 
the virial expansion up to the second virial coefficient. 
In Sec.3 we show our results for the state equation 
of the real pion gas, i.e. we 
obtain the pressure versus the density and the temperature. In this 
case the second virial coefficient is computed both using 
the phase shifts 
that are obtained from standard ChPT as well as those using 
an Inverse Amplitude Method (IAM) fit to pion data up to about
1 GeV. In Sec.4 we obtain the 
quark condensate also in terms of the temperature and the density. 
This will make possible to obtain an approximate phase diagram for 
the pion gas. Finally, in Sec.5 we list the main 
conclusions of this work.

\section{Virial expansion}

\subsection{Generalities}

Let us start by considering a pion gas consisting of $g$ different pion 
 species (for example $g=3$ for the 
$SU(2)_L \times SU(2)_R$ case). The pressure virial expansion 
can be written as \cite{DAS}
\begin{equation}
P=g\,T\left(\frac{M_{\pi}T}{2\pi}\right)^{3/2}\sum_{k=1}^{\infty}B_k(T)
e^{\beta(\mu-M_{\pi})k}=
\frac{q\,T}{\lambda^3}\sum^{\infty}_{k=1}B_k(T) \xi^k
\label{pvir}
\end{equation}
where, as usual, $\beta=1/T$, $M_{\pi}$ 
is the pion mass, $\lambda=(2\pi/M_{\pi}T)^{1/2}$ is the thermal 
de Broglie wavelength and, according to our 
previous discussion, $\mu$ is the chemical potential 
associated with the total number of pions. 
Note that the expansion parameter is the fugacity
$\xi=\exp[(\mu-M_{\pi})/T]$.  
As is well known, the
limit $\xi\ll 1$ corresponds to the low-density regime  where the 
virial expansion is expected to work. The density $n$ 
of the pion gas, i.e. the number $n$ of pions per unit of 
volume can be obtained from 
\begin{equation}
n=\left(\frac{\partial P}{\partial \mu} \right)_{V,T}
\label{nfromp}
\end{equation}
The state equation of the system can be obtained by 
replacing the chemical potential in Eq.\ref{pvir} by its
expression in terms of temperature and density that we can obtain from
Eq.\ref{nfromp}.

\subsection{The free-pion gas}

In order to illustrate the above formalism we will consider first 
the free-pion gas. This will be interesting for different reasons: 
first it is the simplest case and, as far as it 
can be treated exactly, it will 
provide a very nice test for the virial expansion. In particular 
we will be able to estimate the range of temperatures and densities
 at which it can 
be safely applied up to the second virial coefficient.
Second, due to the Weinberg low-energy theorems, pions are weakly 
interacting at low energies and thus the free pion gas can be a 
reasonable approximation to the real pion gas in such regime. 
Finally, let us remember that a free boson gas can suffer a 
Bose-Einstein condensation at very low-temperatures. Apart from the
interest that this phase has by itself, we will see that it basically
determines the applicability region of the virial expansion.

Thus, the free-pion gas pressure can be written as
\begin{equation}
P=-\frac{g\, T}{2\pi^2}\int^{\infty}_{0}dp\; p^2
\log\left[1-e^{\beta(\mu-E(p))}\right]
\label{pfree}
\end{equation}
where $E(p)=\sqrt{p^2-M_{\pi}^2}$ is the energy of a relativistic 
pion in terms of the pion momentum, $p$. The above equation can be 
expanded in powers of the fugacity as 
\begin{eqnarray}
P & = & -\frac{g\, T}{2\pi^2}\int^{\infty}_{0}dp\; p^2
\log\left[1-\xi e^{-\beta(E(p)-M_{\pi})}\right] \nonumber \\
& = & \frac{g\, T}{2\pi^2}\int^{\infty}_{0}dp\; p^2
\sum_{k=1}^{\infty}\xi^k \frac{e^{-k \beta (E(p)-M_{\pi})}}{k}
\end{eqnarray}
Hence, comparing with Eq.\ref{pvir}, we get
\begin{equation}
B_k^{(0)}(T)=\frac{g}{2\pi^2}\left(\frac{M_{\pi}T}{2\pi}\right)^{-3/2}
\frac{1}{k}
\int^{\infty}_{0}dp\; p^2
 e^{- k \beta (E(p)-M_{\pi})}
\end{equation}
where the superscript $(0)$ indicates that we are referring to the 
free gas case. Then the first two virial coefficients are given by
\begin{eqnarray}
B_1^{(0)}(T)&=&\frac{g}{(M_{\pi}T)^{3/2}}\sqrt{\frac{2}{\pi}}
\int^{\infty}_{0}dp\;p^2
 e^{-\beta (E(p)-M_{\pi})}                             \nonumber \\
B_2^{(0)}(T)&=&\frac{g}{2(M_{\pi}T)^{3/2}}\sqrt{\frac{2}{\pi}}
\int^{\infty}_{0}dp\;p^2
 e^{-2\beta (E(p)-M_{\pi})}
\label{virfree}       
\end{eqnarray}
Notice that, in the 
low-temperature limit, they satisfy
\begin{equation}
B_k^{(0)}(T)\simeq g k^{-5/2}+ O\left(\frac{T}{M_{\pi}}\right)
\end{equation}
By using Eq.2 and Eq.4 we can obtain the density
\begin{equation}
n=\frac{g}{2\pi^2}\int_{0}^{\infty}dp\;p^2
\frac{1}{e^{-\beta(\mu-E(p))}-1}   
\end{equation}
Now we can study the free-pion gas, both using the exact results, or the
(second order) virial expansion and then compare 
the results. In that way we expect to get an estimate of the 
applicability range of the virial expansion.

First of all, it is intuitively clear that we need $\xi\ll 1$. Therefore,
we cannot use the virial expansion at $\mu=M_\pi$, which  is precisely  
when Bose-Einstein condensation occurs. Let us briefly study this regime.

\subsubsection{Bose-Einstein condensation}
      
In Eq.\ref{pfree}  the ground state contribution  was not 
taken into account, although
it is well known that it plays an essential 
role at very low temperatures. Indeed, Eq.\ref{pfree} should
be rewritten as      
\begin{equation}
P=-\frac{g\,T}{2\pi^2}\int^{\infty}_{0}dp\;p^2
\log\left[1-e^{\beta(\mu-E(p))}\right]
-\frac{T}{V}\log\left[1-e^{\beta(\mu-M_{\pi})}\right]
\label{pBE}
\end{equation}
and the density as
\begin{equation}
n=\frac{g}{2\pi^2}\int_{0}^{\infty}dp\;p^2
\frac{1}{e^{-\beta(\mu-E(p))}-1}+\frac{1}{V}
\frac{1}{e^{-\beta(\mu-M_{\pi})}-1}   
\label{nBE}
\end{equation} 
where the last term is the number of pions with zero momentum per 
unit of volume, which form the so called Bose-Einstein condensate.
We thus define
\begin{equation}
n_0=\frac{1}{V}\frac{1}{e^{-\beta(\mu-M_{\pi})}-1}  
\label{n0}
\end{equation}
Let us now remember that for Eqs.\ref{pBE} and \ref{nBE} to make sense, 
the chemical potential should satisfy $\mu\le M_\pi$.
As a consequence, in the thermodynamic limit, 
where $N,V\rightarrow\infty$ with $N/V$ constant,
we find two phases: If $\mu\le M_\pi$, $n_0=0$. But  
if we lower the temperature, keeping a fixed density,
$\mu$ increases until $\mu =M_{\pi}$. At that point $n_0\neq 0$
and the ground state density starts to grow, forming the Bose-Einstein
condensate. Eventually, at $T=0$, all  pions are in 
the ground state so that $n_0=n$. The critical temperature $T_c$, 
where the phase transition occurs, can be obtained 
numerically from Eq.\ref{n0}.

\begin{figure}
\begin{center}
\hbox{
\psfig{file=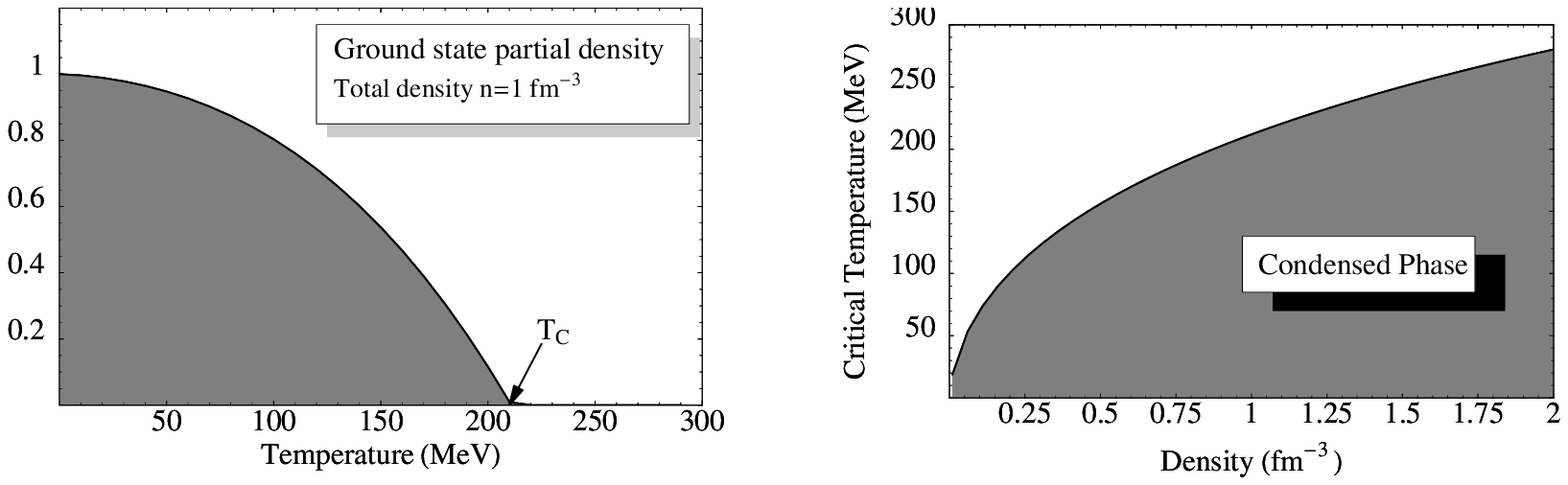,width=6.1in}}

\begin{minipage}{6in}
{\footnotesize {\bf Figure 1.-}  Bose-Einstein condensation in the free-pion
gas: a) Evolution of the 
ground state partial density ($n_0/n$) with the temperature for 
a total density of $n=1$ pion per cubic fermi. b) Phase diagram. 
We plot the critical temperature as a function of the density.}
\end{minipage}
\end{center}
\end{figure}

In Fig.1a we show the ground state partial density, 
defined as $n_0/n$, versus the temperature. For illustrative 
purposes we have chosen a total density $n=1\;fm^{-3}$. 
By using the Landau theory of phase transitions \cite{LAN},
it has been shown that in our case they are 
of second order \cite{KA}. 

From the above discussion it is clear that $T_c$ 
depends on the density. Using Eq.\ref{n0}, we have plotted in Fig.1b
the critical temperature versus the density. That is the Bose-Einstein 
condensation phase diagram of the pion gas. Let us then recall that
in the condensed phase, $\mu=M_\pi$, which means that $\xi=1$, and 
therefore, we {\it cannot} apply the virial expansion.

\subsection{Virial expansion applicability range}

In practice, we will be 
using the second order virial expansion, and we would like to know
where it yields an accurate result. We have just seen that, 
simply because $\xi=1$, there is a range of temperatures and 
densities where we cannot use the virial approach. But the expansion
could still be poor for values of $\xi<1$, that is, $\mu<M_\pi$.
In Fig.2a we show, in the $(T,\mu)$ plane, the error 
in the pressure obtained with the second order virial expansion compared
with the exact calculation using Eq.\ref{pfree}. Notice that the error is
less than $5\%$ if we keep 
$\mu\le 135 \mbox{MeV}$ and 
temperatures lower than 10 MeV.
A similar plot can be obtained for the density.

\begin{figure}
\begin{center}
\hbox{
\psfig{file=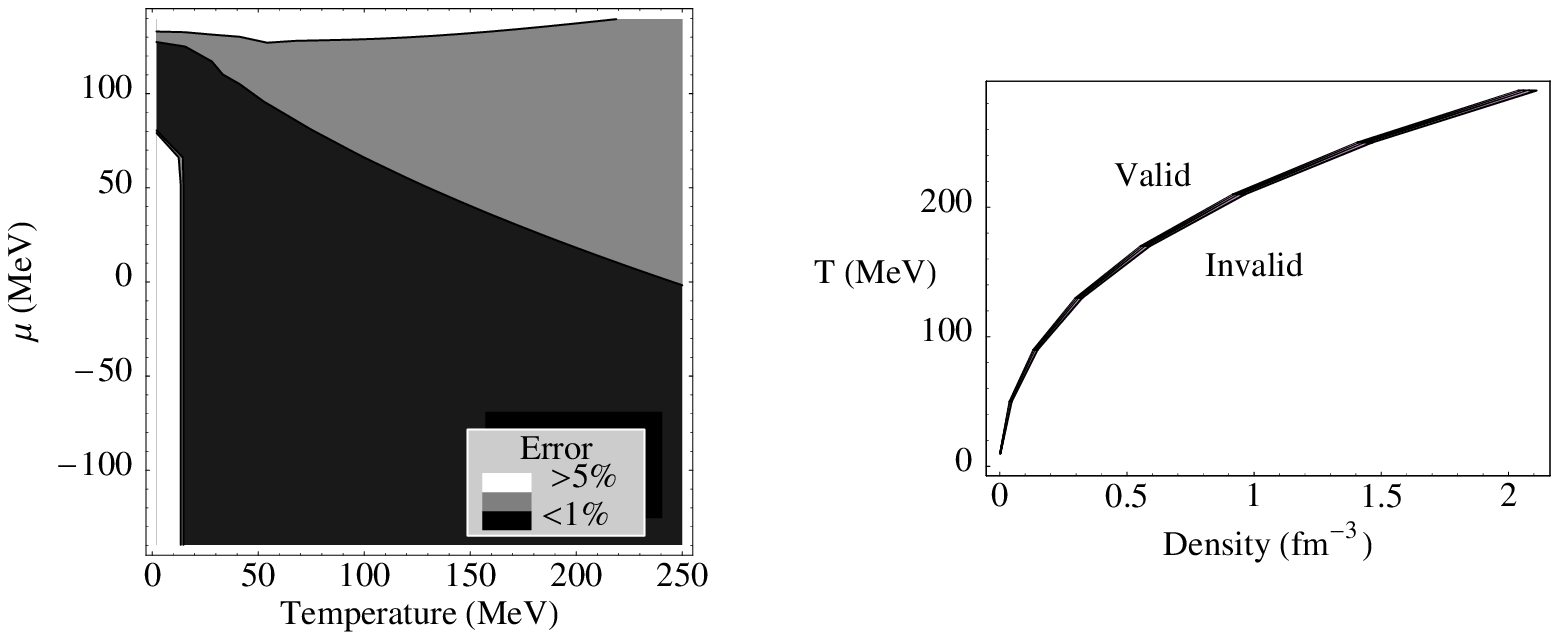,width=6.1in}}

\vskip .3cm

\begin{minipage}{6in}
{\footnotesize  {\bf Figure 2.-} Second order virial expansion 
applicability range: a) In the $(T,\mu)$ plane, within the black area 
corresponds to points where the virial expansion of the pressure
has less than $1\%$ error, the grey area 
stands for less than $5\%$ error, whereas white is more than $5\%$.
b) Applicability region of the virial approach in the $(T,n)$ plane
and the density  (defined as less than $5\%$ error with 
respect to the exact calculation).}
\end{minipage}
\end{center}
\end{figure}

We will see in the following sections that the uncertainties in the
interacting pion gas are of the order of $10\%$. Hence, 
throughout this work we are considering an error 
smaller than  $5\%$ error, both in the 
pressure and the density, as a valid approximation.
Following that criteria, we plot in Fig.2b the estimated
validity region of the second order virial approach. Notice that
 the invalid
region is basically that where Bose-Einstein condensation occurs
(compare Fig.2b with Fig.1b).

\section{The state equation}

In this section we will apply the virial expansion to
the interacting pion gas. Since the second virial 
coefficient is the first one sensible to the interactions 
we have to keep, at least, up to the second 
term in the virial expansion.
That contribution
can be written in terms of the $\pi\pi\rightarrow\pi\pi$ 
elastic scattering phase shifts  
 $\delta_{IJ} (E)$ as \cite{DAS}
\begin{equation}
B_2(T)=B_2^{(0)}(T)+\frac{4e^{2M_{\pi}/T}}{(2\pi M_{\pi}T)^{3/2}}
\int^{\infty}_{2M_{\pi}}
dE E^2 K_1(E/T)\left(\sum_{I,J}(2I+1)(2J+1)\delta_{IJ}(E)\right)
\label{Bint}
\end{equation}
where $K_1$ is the modified Bessel function with asymptotic behavior
\begin{equation}
K_1(x)\simeq \sqrt{\frac{\pi}{2x}}e^{-x}
\end{equation}
for large $x$.

It is important to notice
that in Eq.\ref{Bint} 
we could simply use the experimental phase shifts, without
any reference to any underlying physical theory. In this way
we can compute
quantities as the pressure or the density,  which
is enough to obtain the state equation.  But we could not
calculate their derivatives with respect to, for instance, the pion
mass, which is needed to study the $\langle \bar{q}q\rangle$ condensate.

Therefore, in this work we will obtain the phase shifts
with two different approaches: On the one hand, we will use
ChPT to $O(p^4)$, with the parameters proposed in \cite{GaLe,RiDo}
which fits the experimental data 
on elastic pion scattering for the $(I,J)=(0,0),(1,1),(2,0)$
channels, up to energies of the order of 0.5 GeV. 
The interest of using ChPT relies on the fact that it yields
a systematic expansion in terms of the external momenta and 
masses of the pions. That will be extremely relevant 
in order to obtain other thermodynamic properties that 
may require derivatives with respect to the masses. That is indeed
the case of the chiral condensate, that will be treated 
in the next section.
On the other hand, it has been shown \cite{IAM}
that the ChPT fits to pion scattering can be extended to higher energies
using the Inverse Amplitude Method (IAM). Indeed, it yields
remarkably good fits up to approximately 1 GeV. That could 
help to obtain 
a better estimate of the pressure, and the equation of state,
although maybe not of its derivative with respect to the pion mass. 

Moreover, as far as we will
be interested in temperatures $T\leq 300\, \mbox{MeV}$, 
the Bessel function 
in Eq.\ref{Bint} will suppress the $\delta(E)$ contributions 
at high energies. Thus we can also use the IAM phase shifts
to estimate how big is the error on the pure ChPT
results due to these higher energy contributions.
However, note that, as far as both  methods 
reproduce pretty well the experimental data at low energies, 
our results concerning the state equation of the pion gas a low 
temperatures can be considered as quite realistic independently 
of the fact that we have used ChPT or the IAM.

Thus, from the phase-shifts coming from standard
ChPT and from ChPT complemented with the IAM method, 
together with Eqs.\ref{virfree} and
\ref{Bint} we can obtain the
second virial coefficient. Using Eq.\ref{pvir} we can therefore
compute the value of the pressure as a
function of $T$ and $\mu$. We show the results in Fig.3.
In particular, we plot the pressure dependence on 
the temperature for $\mu=0$, which is well within the
applicability range of the virial approach
(This case is interesting since it corresponds to the 
canonical ensemble \cite{GELE}). 
One of the three curves is the free-pion gas and the other two  
the interacting case both using standard ChPT or ChPT with the IAM.

\begin{figure}

\begin{center}
\hspace*{.5cm}
\hbox{
\psfig{file=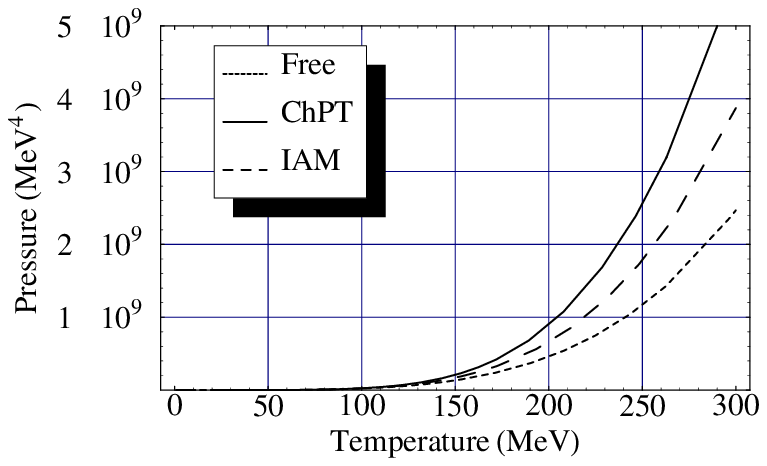,width=3.5in}
}

\vskip .2 cm

\begin{minipage}{6in}
{\footnotesize  {\bf Figure 3.-} The pressure of an
interacting pion gas as a function of the temperature.
We plot the curves both for the phase shifts obtained
from pure ChPT as well as with the IAM, together with
that of the free pion gas, as a reference.
}
\end{minipage}
\end{center}
\end{figure}

Moreover, by solving numerically Eq.\ref{nfromp} we can obtain $\mu$ 
in terms of the density $n$ and $T$. Thus we can finally 
compute the pressure as a function of the temperature and 
the density, $P=P(T,n)$, i.e. we obtain the state equation 
for a pion gas with realistic interactions. In Figs.4a and 4b 
we show a three dimensional plot of the state equation surface, 
for standard ChPT and the IAM, respectively. As we have already discussed
in the previous section, there are parts of the $(T,n)$ 
space that we cannot
explore within the virial approach and have been left blank. Out of that
area, we can trust the virial expansion. Nevertheless we expect to
find some differences depending on whether we complement ChPT with
the IAM or not, which should become bigger at higher temperatures.
Nevertheless,
it can be noticed that the numerical difference between Fig. 4a and 4b,
 is of the order of $5\%$ at $T=300$ MeV and 
$n=2\, \hbox{fm}^{-3}$. As we are going to be interested in
temperatures below 300 MeV, it seems that standard ChPT yields a state 
equation with sufficient accuracy.

\begin{figure}
\begin{center}
\hbox{
\psfig{file=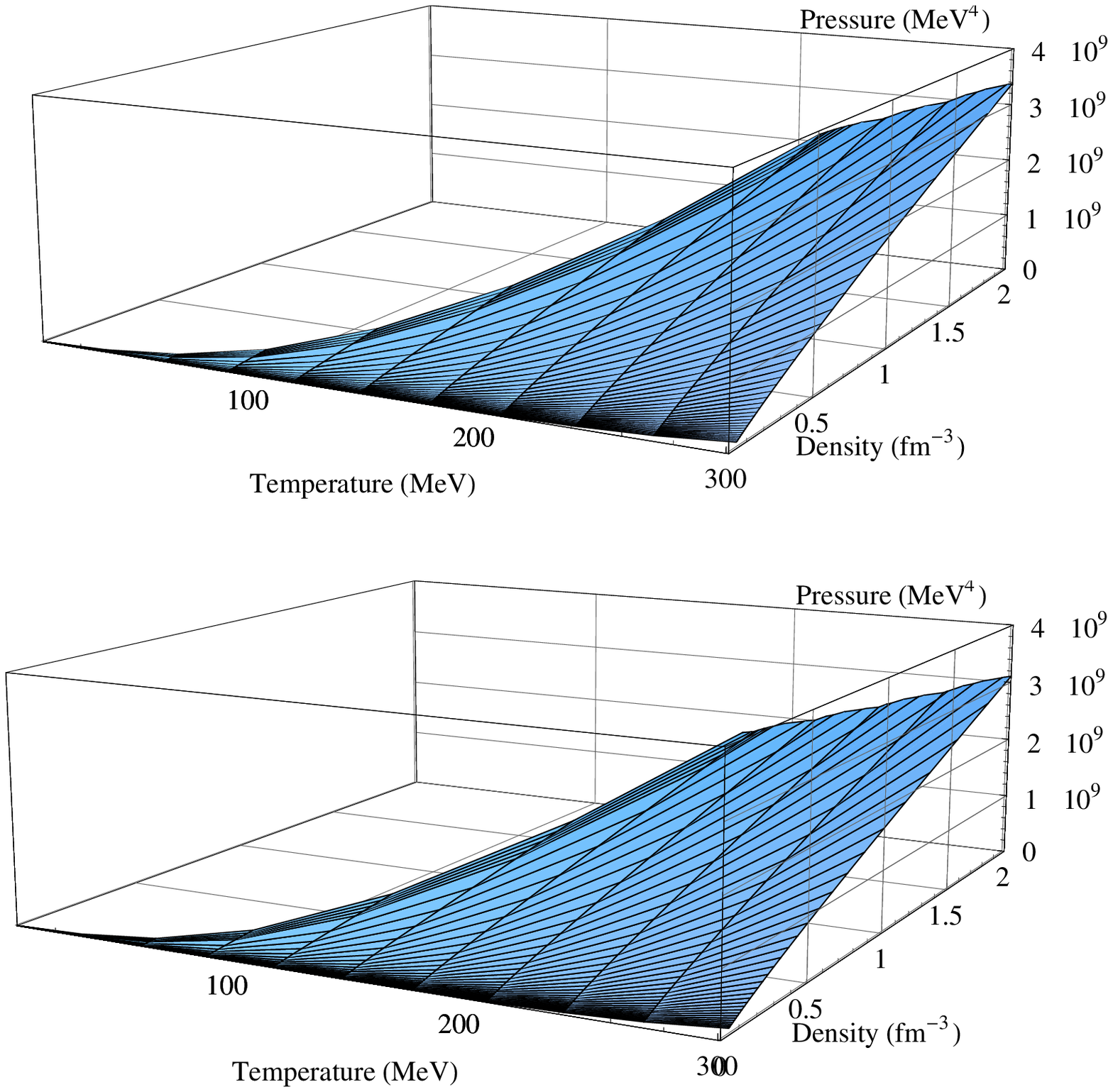,width=6.1in}
}

\vskip 1.5 cm

\begin{minipage}{6in}
{\footnotesize   {\bf Figure 4.-} State equation for an interacting
pion gas. The blank area is out of the applicability
reach of the virial approximation.
We plot the pressure as a function of temperature and density:
a) using standard ChPT phase shifts. 
b) complementing ChPT with the IAM.
}
\end{minipage}
\end{center}
\end{figure}

Let us remark that we are strictly using an $SU(2)$ model 
and therefore we do not have other light particles, like kaons or etas, 
present.
In a realistic gas, the effect of these other particles should also
be taken into account, although it is suppressed by Boltzmann factors 
(see \cite{GELE}). Nevertheless, it is also possible to apply the 
virial formalism to 
a gas with different species. The phase shifts can also be obtained from a
generalization of the IAM \cite{prl}, which fits the existing data up
to 1.2 GeV and provides predictions when data are not available.
Such calculations could indicate the time scale at which the approximation
of pion number conservation is valid.
However, although the present approach can be generalized
to include such extensions, they are beyond our present scope.
 
Once we have delimited the applicability constraints of standard ChPT 
with the virial expansion, we will use it in the next section to study
the chiral transition and the quark condensate.

\section{The quark condensate}

One of the topics of hadronic physics which has not completely
been settled concerns the chiral phase transition. At some critical 
temperature it is expected that the spontaneously broken 
chiral symmetry is restored giving rise to a new phase. The 
nature of that phase is not known but it has been 
argued that the transition itself
 is of second order for the $SU(2)_L \times 
SU(2)_R$ case \cite{WIL}, at least in the chiral limit, or 
very soft first order, i.e. with a very small latent heat 
\cite{LE}. 

However, since the broken phase occurs at low
temperatures, it should be mostly made out of pions.
Hence, a better understanding of the
pion gas may be a useful approach
to study the chiral transition.
Indeed, the broken phase is characterized by 
a non-vanishing order parameter, which 
is usually identified with the quark condensate 
$\langle\bar q q\rangle_T$. 
Following \cite{GELE}, it can be written as
\begin{equation}
\frac{\langle \bar q q\rangle_T}
{\langle \bar q q \rangle_0}
=\left(1 + \frac{c}{2M_\pi F^2}\frac{\partial P}{\partial M_\pi} \right)
\label{cucu}
\end{equation}  
where $F$ is basically the pion decay constant without $O(p^4)$
corrections, m is the $u$ and $d$ averaged mass,  and
\begin{equation}
c=-\frac{F^2}{\langle \bar q q\rangle_0}
\frac{\partial M^2_\pi}{\partial m}
\end{equation}
For our calculations we have taken the numerical values considered 
in \cite{GELE}, namely, $c=0.90\pm0.05$ and $F=88.3\pm 1.1$ MeV.

In order to compute the quark condensate, and following the
arguments of previous sections, we will only
use the standard ChPT phase shifts. 
They seem accurate enough 
for our purposes up to the relevant temperatures and, 
at the same time, they have the appropriate $M_{\pi}$ 
dependence, which is essential in Eq.\ref{cucu}. 

\begin{figure}
\begin{center}

\hspace*{.1cm}\hbox{
\psfig{file=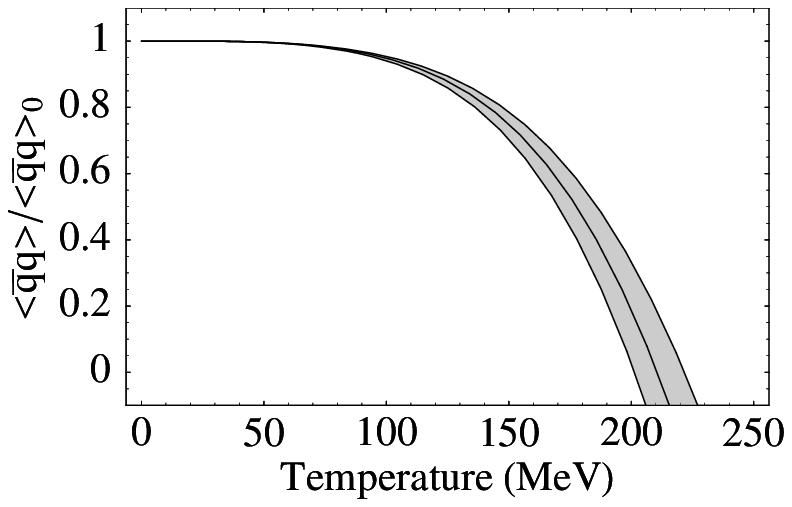,width=3.5in}
}

\vskip .3cm

\begin{minipage}{6in}
{\footnotesize   {\bf Figure 5.-} Evolution of
the chiral condensate with the temperature at $\mu=0$. The shaded area
covers the uncertainties in the different parameters, that have 
been explained in the text.}
\end{minipage}
\end{center}
\end{figure}

In Fig.5 we show the evolution of the chiral condensate when
$\mu=0$, for standard ChPT. The shaded area covers the uncertainties
in $F$ in $c$ as well as those due to the chiral 
parameters $\bar{l}_1=-0.62\pm0.94$, $\bar{l}_2=6.28\pm0.48$ 
(see \cite{RiDo}), $\bar{l}_3=2.9\pm2.4$ and 
$\bar{l}_4=4.3\pm0.9$ (see \cite{GaLe}). We have also included 
in the uncertainties the effect of choosing $M\pi$ as that of the
charged or the neutral pions, although that effect is rather small.

Let us remember once more that the $\mu=0$ results are formally equivalent
to those obtained in the canonical ensemble (see \cite{GELE}). 
The critical 
temperature can be estimated as the point where the condensate 
vanishes, and it is around 220 MeV.
These numbers should be interpreted extremely carefully, since
the temperatures are quite high. Nevertheless the plots seem to indicate a
clear tendency towards chiral symmetry restoration above 200 MeV.
Our curves continue down to negative values, but at that point
the system should be in the unbroken phase and our formalism 
is no longer appropriate. The results are in a remarkably good agreement
with those of \cite{GELE}, which were obtained from a full 
three loop ChPT calculation. This fact is a nice
check of our calculations and provides further 
support for their conclusions.
 
In order to see the effects of density on the chiral 
phase transition, we show in Fig.6a the value of the 
chiral condensate in the $(T,\mu)$ plane. The region where the
virial expansion is not applicable, $\mu\geq 135$ MeV, which is located
at the far end of the picture, is almost imperceptible due to the scale.
The points where  $\langle\bar{q}q\rangle/\langle\bar{q}q\rangle_0=0$
are an estimate of the critical temperature 
given a fixed chemical potential.
The results suggest that there are values of the chemical 
potential for which there is no chiral phase transition. 

\begin{figure}
\begin{center}
\hspace*{.1cm}\hbox{
\psfig{file=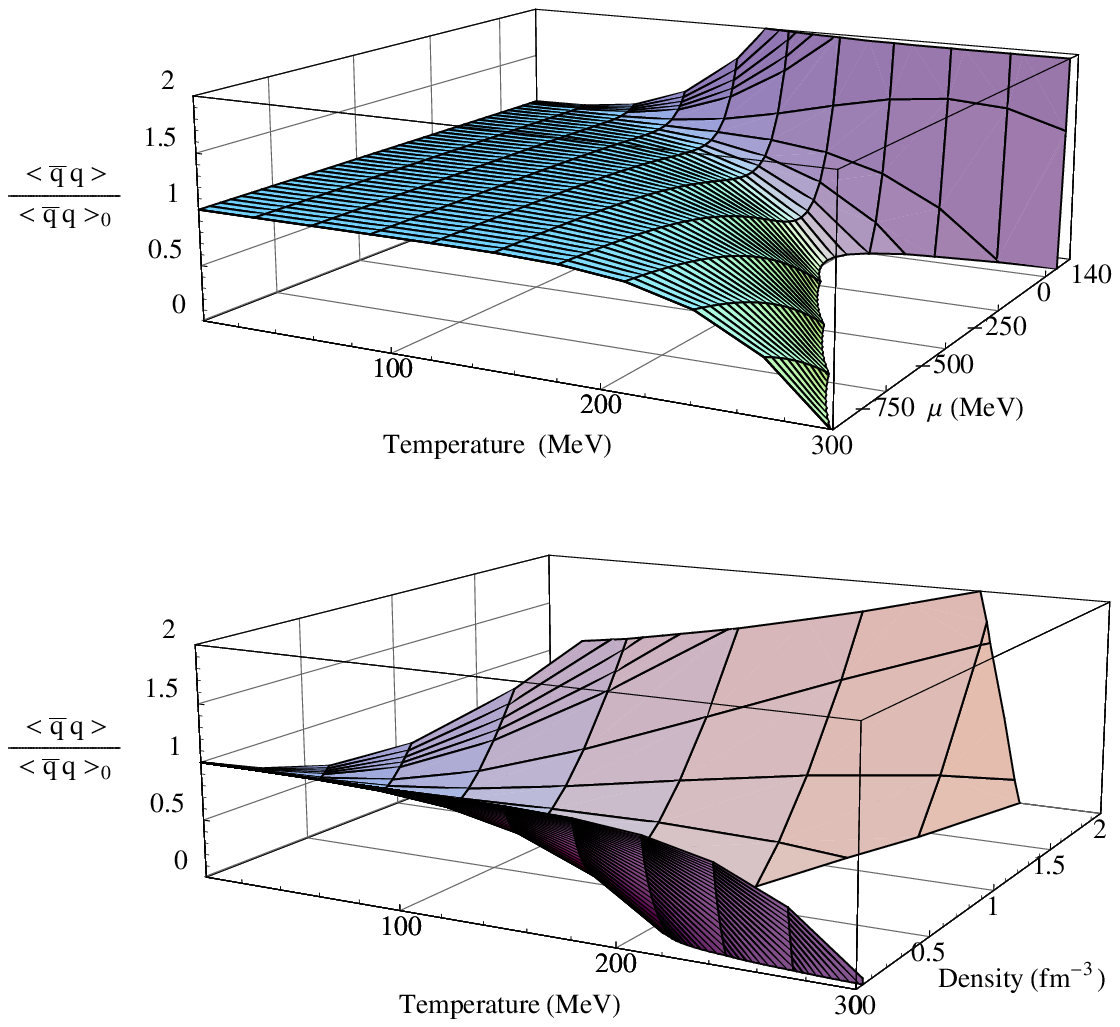,width=6.1in}}

\vspace{.5cm}

\begin{minipage}{6in}
{\footnotesize   {\bf Figure 6.-} a) The value of
 $\langle\bar{q}q\rangle/\langle\bar{q}q\rangle_0$ 
as a function of $T$ and $\mu$. b) The value of
 $\langle\bar{q}q\rangle/\langle\bar{q}q\rangle_0$ as a function of the 
temperature and the density.}
\end{minipage}
\end{center}
\end{figure}

From now on we will also give results just for the central values
 of the different parameters and $M_\pi=139.57$.

Following the above reasoning it is therefore possible to obtain a 
qualitative phase diagram, which we have plotted in Fig.7a. Again, there seem
to be limits, both at high and low chemical potential, where there is no 
phase transition (namely, $T_C\rightarrow\infty$). 
Moreover, for a given temperature, it seems possible,
by decreasing $\mu$, to change from the broken to the unbroken phase, 
and back to the broken phase again.

\begin{figure}
\begin{center}
\hspace*{.1cm}\hbox{
\psfig{file=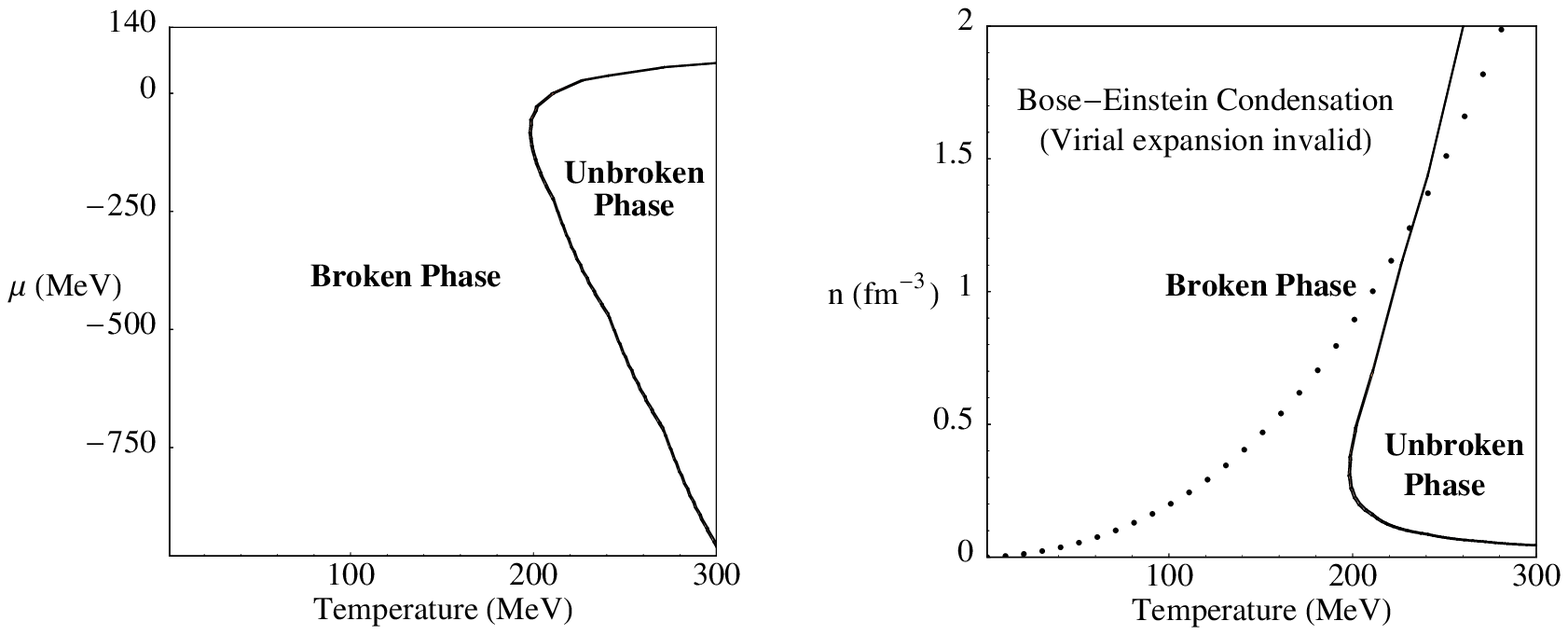,width=6.1in}
}

\vspace{.5cm}

\begin{minipage}{6in}
{\footnotesize  {\bf Figure 7.-} a) Phase diagram of the chiral 
transition in the $(T,\mu)$ plane. b) Phase diagram of 
the chiral transition in the
$(T,n)$ plane. The continuous line separates the phases with
broken and unbroken chiral symmetry. The phase with 
Bose-Einstein condensation is located above the dotted line.
}
\end{minipage}

\end{center}
\end{figure}

Apparently there are several competing effects acting on the condensate. 
On the one hand, the temperature tends to destroy any correlation. 
On the other hand, the density tends to restore the broken phase if it
is very high, since then the interactions of the pions can overcome 
the effect of the temperature  (at these energies, each microscopic scattering
follows the constraints of chiral symmetry breaking).
But that also happens at lower densities, since the collisions are so
scarce that the temperature is not able to break any existing correlation.
This last effect can be easily understood since the low
density limit should lead to the usual pion scattering at $T=0$.
But that would be nothing but ChPT, which incorporates in its own definitions
the chiral symmetry breaking.

The plots in terms of the chemical potential have the advantage that the 
non-applicability region is easily defined and localized. 
In addition, it only
occupies a tiny region of the interesting parameter space. However,
the interpretation is less intuitive.

Therefore, we show in Fig.6b the values of the chiral condensate
as a function of the temperature and the density. Note that now we are using
the virial expansion both for  the condensate itself (since it
is obtained from the pressure) {\em and the density}. 
Therefore, there  are parts of the parameter
space where we cannot apply the virial approach and have been intentionally
left blank.  From those results, we can extract once more the phase 
diagram, this time in the $(T,n)$ plane. We have plotted it in Fig.7b.
Notice that the area within the dotted line is out of  reach for
our virial approach, although it seems plausible that it belongs
to the broken phase. Nevertheless,
within the temperature and density region that we can explore,
we see how the critical temperature grows extremely rapidly
when we decrease the density, tending to infinity at $n\rightarrow0$.
In contrast,
it is not clear whether that is also the case at high densities. 
From the phase diagram we can
see that the critical temperature grows very softly with the 
density above $0.5 \, fm^{-3}$, but we soon cross to the
region where the virial approach is not applicable.

\section{Conclusions}

In this work we have studied the 
thermodynamic properties 
of a hadronic gas at low temperature and density in the
absence of baryons. In such a case the hadronic matter 
can be understood as basically a gas of pions.  

In order to describe the low-energy pion dynamics we have
taken into account two main facts: First, that the spontaneous 
chiral symmetry breaking makes possible the 
use of Chiral Perturbation Theory (ChPT) to
describe $\pi\pi$ elastic scattering, which fits 
rather well the low-energy data up to around $0.5$ MeV. 
Using scattering
amplitudes unitarized with the Inverse Amplitude
Method, which fit the data almost up
to 1 GeV, we have indeed checked that, as far as $\pi\pi$
scattering is concerned, the standard ChPT approach
is rather accurate in the region of interest, below 
the chiral phase transition.

Second it is a matter of fact that 
elastic unitarity is fulfilled up to around 1 GeV.
This allows us  to consider the total pion number 
as an approximate conserved quantity at low temperatures. 
Hence, we can define the corresponding chemical 
potential and the grand canonical ensemble.  

With the above considerations, we have chosen a
formalism where it 
is possible to derive the thermodynamic functions directly 
from the $S$ matrix (phase shifts) by means of the virial 
expansion. In order to find the
region where the virial expansion yields an accurate result,
we have first used the free gas case, for which there are
closed expressions. That has also allowed us to study
the Bose-Einstein condensation, which, apart from the
interest in itself, basically determines the region where the
virial approach is not applicable. Indeed, for a given density,
if we want to use the virial expansion, we have to
be several MeV above the Bose-Einstein phase transition.

Next, we have included the interactions and obtained the 
state equation $P=P(T,n)$, both for the standard ChPT and
unitarized scattering amplitudes. The agreement seems to be
good enough to keep simply the standard description.
Finally, using the interacting gas free energy  
we have also studied the chiral condensate dependence
on the temperature and the density. Our results
for zero chemical potential are in very good agreement
with previous calculations using a different approach.
However, by considering density effects we have now
been able to obtain phase diagrams and to study the
interplay of temperature, which favors the melting
of the chiral condensate, with the density, which tends to
enhance the chiral symmetry breaking built in the $\pi\pi$
low energy interactions. From our phase diagrams we can
learn how
by diluting the hadronic gas it is possible to raise the critical
temperature. At high densities, there also seems to be a very soft
increase of the critical temperature, but that interpretation is more
subtle due to the breaking of the virial expansion.
Of course, more detailed calculations should also
consider heavier particles, like kaons or etas within
the $SU(3)_L \times SU(3)_R$ chiral scheme, as well as baryons.
These contributions grow with the temperature although, 
in general, they are expected to lower the critical 
temperature \cite{GELE}.

We would finally like to stress the phenomenological nature
of our work since
the only ingredients are chiral symmetry, unitarity, 
the virial expansion and the pion phase shifts.

\section*{Acknowledgments}

J.R.P. would like to thank the Theory Group at SLAC for their 
kind hospitality and the Spanish
Ministerio de Educaci\'on y Cultura for a Fellowship. This work 
has been partially supported by the 
Spanish CICYT under contract AEN93-0776.

\end{document}